\documentclass[trackchanges]{aastex701}

\begin{document}

\title{X-Ray Observation of Type II Supernova 2024ggi}

\author[orcid=0009-0006-1696-2118]
{Elisa J. Gao}
\affiliation{University of Chicago, Department of Astronomy and Astrophysics}
\email[show]{elisagjy@uchicago.edu}  

\author[orcid=0000-0000-0000-0002]{Vikram V. Dwarkadas} 
\affiliation{University of Chicago, Department of Astronomy and Astrophysics}
\email{vikramvd@uchicago.edu}

 \begin{abstract}

 We investigate the \textit{Chandra} and \textit{XMM-Newton} X-ray emission from the core-collapse SN 2024ggi. Spectral fitting is carried out using the vapec model. We find X-ray temperatures between 5-7 keV, lower than the value assumed by previous authors by about a factor of 4-6. However the unabsorbed flux is consistent with published values within the error bars. The X-ray luminosity evolution is somewhat steeper than expected for a steady wind, although it could be consistent with a steady wind within the error bars.

\end{abstract}




\section{Introduction} 

Type II supernovae are explosions of massive stars ($\gtrsim 8,M_\odot$). These stars undergo gravitational collapse at the end of their lives, ejecting most of the outer envelope. The resulting shock wave(s) generated by the explosion accelerate the ejected material to velocities exceeding 1000 times that of the pre-explosion stellar wind. The shocks heat up the plasma to temperatures exceeding 10$^6$ K, resulting in X-ray emission. 

Supernova 2024ggi was discovered on April 11, 2024, by the ATLAS instrument, at UTC 03:22:35.616 (JD=2460411.64069), with RA 11:18:22.087 and DEC -32:50:15.27. It was classified as a Type IIP supernova \citep{Srivastav,Zhai}, one that shows a plateau in its optical light curve. Exploding about a year after the well-known SN 2023ixf, it was also at a relatively close distance of 7.2 $\pm$ 2.2 Mpc \citep{sahaetal06}, and therefore extensively observed by a variety of telescopes at all different wavelengths.

\citet{Lutovinov} reported \textit{SRG/ART-XC} observations obtained shortly after the discovery, with fluxes of $(3.0 \pm 0.3)\times10^{-13}$ erg cm$^{-2}$ s$^{-1}$ in the 4--12 keV energy band and $(1.9 \pm 0.5)\times10^{-13}$ erg cm$^{-2}$ s$^{-1}$ in the 12--20 keV energy band, respectively. Given that the flux in the lower-energy band exceeds that in the higher-energy band, the electron temperature probably peaks in the lower energy band, consistent with what we find below. 

\citet{Ferdinand_2026} applied an absorbed thermal bremsstrahlung model to analyze the X-ray light curve of SN 2024ggi during the first 85 days after the explosion using the \textit{XSNAP} pipeline, which they developed. They were unable to constrain the model temperatures, and instead assumed fixed values of the temperature $kT$ that had been derived from an analysis of SN 2023ixf \citep{chandraetal24}. These values assume apriori that the SN is expanding in a steady wind, the expansion is self-similar, and that the ejecta density is decreasing as ${\rho}_{ej} \propto v^{-10}$. These assumptions have not been validated for SN 2024ggi. In the case of SN 2024ggi, they resulted in temperatures $>$ 30 keV for the \textit{Chandra} observations, and $>$ 20 keV for the XMM observations. \citet{Ferdinand_2026} reported a steady mass-loss rate for the progenitor star of $(6.2 \pm 0.2) \times 10^{-5}$ $M_\odot$ yr$^{-1}$, assuming a wind velocity of $v_{\rm wind} = 20$ km s$^{-1}$. We note that this is about 200 times less than that of SN 2023ixf in the first 10 days or so, and about a factor of 2 lower at later times \citep{jacobsongalan25a, jacobsongalanetal25b}. Since the mass-loss rates of the material into which the SNe are expanding are different, it is likely that the expansion velocities are also different. The post-shock temperature, being proportional to  the expansion velocity, would likely be different. Therefore, using a temperature function derived for SN 2023ixf does not seem optimal for SN 2024ggi.

In this research note, we re-examine the X-ray observations of SN 2024ggi. We aim to determine the best-fit temperatures from the X-ray spectra, and compute the resultant X-ray light curve of SN 2024ggi.

\section{Observations} 

Table 1 summarizes the observation parameters as well as the spectral fits.

\subsection{Chandra}
We process the two available Chandra observations (Chandra Data Collection DOI: \href{https://doi.org/10.25574/cdc.630}{https://doi.org/10.25574/cdc.630}) using CIAO v4.16 and CALDB v4.11.3. For both epochs, we extract source and background spectra using a circular region with radii of 4$^{\prime\prime}$, which encompasses more than 90\% of the encircled energy. The spectra and corresponding response files are generated using the \textit{specextract} function. Spectral analysis is performed with the \textit{Sherpa} software.

We fit both \textit{power-law} and \textit{vapec} models to the observations, The latter describes thermal emission from collisionally-ionized diffuse gas calculated from the AtomDB atomic database (http://atomdb.org/). We eventually adopted the latter due to the presence of a prominent Fe emission line in the first observation (ObsID 29383). The Galactic column density is assumed to be $0.018 \times 10^{22}\,\mathrm{cm^{-2}}$. An additional column due to material surrounding the SN, or between the SN and the observer, is also considered. The total column density is listed in Table 1. A redshift of 0.0024 was assumed \citep{Koribalski_2004}. 

We are able to obtain the column density ($N_{\rm H}$) and temperature ($kT$) from the fit itself, and do not need to assign a temperature. As noted from Table 1, the derived temperature is lower than that assumed by \citet{Ferdinand_2026}.

\subsection{XMM-Newton}
There are two XMM-Newton observations of SN 2024ggi. We reduced the data from the MOS1 and MOS2 cameras using XMM-SAS v21.0.0. Source and background spectra are extracted from circular regions with radii of 30$^{\prime\prime}$. We again use the \textit{Sherpa} software to fit the observations, using a \textit{vapec} model, which provides the best fit to the \textit{Chandra} spectra.

Due to the low counts, we are unable to constrain the $kT$ value of the first XMM observation from the fit, although we can constrain the temperature in the second \textit{XMM} observation at 85 days. Since the temperature in the Chandra observation at $\sim16$ days and the XMM observation at 85 days are roughly the same, we fix the $kT$ to be 5.5keV keV for the \textit{XMM} observation at 54.5 d. This approach provides reasonable estimates of $N_{\rm H}$ and the unabsorbed flux for the first XMM observation. Raising the temperature to 6 keV decreases the flux by 3\%, a negligible change.

\subsection{Swift}
We then proceed to analyze the Swift data on SN 2024ggi, using the XSELECT package from HEASOFT v6.35.2. Owing to the short exposure times (typically $\lesssim 2$ ks) of all available observations, we combine multiple epochs to improve the signal-to-noise ratio. However, after combining all observations obtained during the first five days following the explosion, we are unable to detect a signal in the combined observation, within a 25$^{\prime\prime}$ radius of the source co-ordinates.  Since the supernova undergoes rapid evolution during the early phases after the explosion, combining observations over a longer time interval would not accurately represent its behavior at a given epoch since the flux is changing rapidly. The SN is growing fainter with time, making it unlikely that it will be detected in subsequent observations. Therefore, we choose not to include Swift data in the subsequent analysis.

\section{Results} 

The $N_{\rm H}$, $kT$, and unabsorbed flux values of SN 2024ggi in the $0.5$--$10$ keV energy range are listed in Table~1. We also include the observation reported by \citet{Lutovinov} using \textit{SRG/ART-XC} shortly after the explosion. Note that the energy range for the SRG observation is different and the temperature was not provided.

\begin{deluxetable*}{lccccc} [hbt!]
\tablecaption{Flux from observations \label{tab:flux}}
\tablehead{
\colhead{Observation ID} & 
\colhead{Days after explosion} & 
\colhead{Instrument} & 
\colhead{$N_H$ ($10^{22}$ cm$^{-2}$)} & 
\colhead{kT (keV)} & 
\colhead{Unabsorbed flux} \\ 
\colhead{} & 
\colhead{} & 
\colhead{} & 
\colhead{} & 
\colhead{} & 
\colhead{($10^{-13}$ erg cm$^{-2}$ s$^{-1}$)}
}
\startdata
 & 1.82 & SRG/ART-XC  &  &  & 3.0 ${\pm 0.3}$ $^{*}$\\
29383 $^{a}$ & 10.55 & Chandra ACIS-S  & 3.97 ${\pm 1.05}$ & 6.55 ${\pm 2.48}$ & 4.85 $^{+1.71}_{-1.35}$ \\
29384 $^{b}$ & 15.85 & Chandra ACIS-S & 1.03 $\pm 0.74$ & 5.44 $\pm 2.10$ & 2.26 $^{+1.34}_{-0.80}$ \\
0882480901  & 54.54 & XMM MOS & 0.018 $^{+0.43}$ & 5.5 $^{c}$& 0.62 $^{+0.34}_{-0.36}$ \\
0882481001 & 85.04 & XMM MOS & 0.018 $^{+ 0.07}$ & 5.6 $\pm 0.44$ & 0.46 $^{+0.13}_{-0.12}$ \\
\enddata
\tablenotetext{*}{\cite{Lutovinov}; 4-12 keV range}
\tablenotetext{a}{Thawed Fe.}
\tablenotetext{b}{Thawed Ca.}
\tablenotetext{c}{Fixed value.}
\end{deluxetable*}

By fitting a power law to the unabsorbed flux obtained from \textit{Chandra} and \textit{XMM-Newton} observations, we find that the flux evolves as $t^{-1.21\pm 0.22}$. Following the procedure described in Section~3.2 of \citet{2011dh}, who performed a similar analysis for the Type II supernova SN~2011dh, we estimate the circumstellar medium (CSM) density profile parameter $s$ (where $\rho_{\rm CSM} \propto r^{-s}$). Exploring ejecta density indices in the range $n = 10-30$, (${\rho}_{\rm ej} \propto v^{-n}$) we find $s \approx 2.1 \pm 0.12$. Within the error bars, this result could be consistent with a steady wind with constant parameters, as derived by \citet{Ferdinand_2026}.

\section{Discussion} 
We analyze the available \textit{Chandra} and \textit{XMM-Newton} observations of SN 2024ggi. Unlike~\citet{Ferdinand_2026}, we do not fix the X-ray temperature, but instead derive it directly from spectral fitting whenever possible. We find temperatures that are lower by a factor of $\sim4$--6 compared to those assumed by \citet{Ferdinand_2026}. This does not significantly affect the X-ray flux. The resulting X-ray flux is generally consistent with the work of \citet{Ferdinand_2026} within the uncertainties. Although the X-ray luminosity declines somewhat faster than what would be expected for a steady-wind ($L_X \propto t^{-1}$), it is consistent with a steady wind within the error bars. Our inferred CSM density profile, characterized by $s \approx 2.1 \pm 0.12$, is also broadly consistent with the steady-wind value of $s = 2$.

\bibliography{citation.bib}

@ARTICLE{Lutovinov,
       author = {{Lutovinov}, A.~A. and {Semena}, A.~N. and {Mereminskiy}, I.~A. and {Sazonov}, S. Yu. and {Molkov}, S.~V. and {Tkachenko}, A. Yu. and {Arefiev}, V.~A.},
        title = "{SRG/ART-XC detects SN2024ggi in X-rays}",
      journal = {The Astronomer's Telegram},
         year = 2024,
        month = apr,
       volume = {16586},
        pages = {1},
       adsurl = {https://ui.adsabs.harvard.edu/abs/2024ATel16586....1L}
}

@ARTICLE{Zhai,
       author = {{Zhai}, Q. and {Li}, L. and {Zhang}, J. and {Wang}, X.},
        title = "{LiONS Transient Classification Report for 2024-04-11}",
      journal = {Transient Name Server Classification Report},
         year = 2024,
        month = apr,
       volume = {2024-1031},
        pages = {1},
       adsurl = {https://ui.adsabs.harvard.edu/abs/2024TNSCR1031....1Z}
}

@ARTICLE{Srivastav,
       author = {{Srivastav}, S. and {Chen}, T.~W. and {Smartt}, S.~J. and {Nicholl}, M. and {Smith}, K.~W. and {Young}, D.~R. and {Fulton}, M. and {McCollum}, M. and {Moore}, T. and {Weston}, J. and {Sheng}, X. and {Aamer}, A. and {Angus}, C.~R. and {Ramsden}, P. and {Shingles}, L. and {Gillanders}, J. and {Rhodes}, L. and {Andersson}, A. and {Stevance}, H. and {Denneau}, L. and {Tonry}, J. and {Weiland}, H. and {Lawrence}, A. and {Siverd}, R. and {Erasmus}, N. and {Koorts}, W. and {Jordan}, A. and {Suc}, V. and {Rest}, A. and {Stubbs}, C. and {Sommer}, J.},
        title = "{ATLAS24fsk (AT2024ggi): discovery of a nearby candidate SN in NGC 3621 at 7 Mpc with a possible progenitor detection}",
      journal = {Transient Name Server AstroNote},
         year = 2024,
        month = apr,
       volume = {100},
        pages = {1},
       adsurl = {https://ui.adsabs.harvard.edu/abs/2024TNSAN.100....1S}
}

@article{Ferdinand_2026,
doi = {10.3847/1538-4357/ae4ec6},
url = {https://doi.org/10.3847/1538-4357/ae4ec6},
year = {2026},
month = {apr},
publisher = {The American Astronomical Society},
volume = {1001},
number = {1},
pages = {26},
author = {Ferdinand and Jacobson-Galán, W. V. and Kasliwal, M. M. and Zimmerman, Erez A.},
title = {XSNAP: An X-Ray Supernova Analysis Pipeline with Application to the Type II Supernova 2024ggi},
journal = {The Astrophysical Journal}
}

@Article{2011dh,
AUTHOR = {Gao, Elisa J. and Dwarkadas, Vikram V.},
TITLE = {Analysis of 14 Years of X-Ray Emission from SN 2011DH},
JOURNAL = {Universe},
VOLUME = {12},
YEAR = {2026},
NUMBER = {1},
ARTICLE-NUMBER = {16},
URL = {https://www.mdpi.com/2218-1997/12/1/16},
ISSN = {2218-1997},
DOI = {10.3390/universe12010016}
}

@ARTICLE{sahaetal06,
       author = {{Saha}, A. and {Thim}, F. and {Tammann}, G.~A. and {Reindl}, B. and {Sandage}, A.},
        title = "{Cepheid Distances to SNe Ia Host Galaxies Based on a Revised Photometric Zero Point of the HST WFPC2 and New PL Relations and Metallicity Corrections}",
      journal = {\apjs},
         year = 2006,
        month = jul,
       volume = {165},
       number = {1},
        pages = {108-137},
          doi = {10.1086/503800},
archivePrefix = {arXiv},
       eprint = {astro-ph/0602572},
 primaryClass = {astro-ph},
       adsurl = {https://ui.adsabs.harvard.edu/abs/2006ApJS..165..108S}
}

@ARTICLE{chandraetal24,
       author = {{Chandra}, Poonam and {Chevalier}, Roger A. and {Maeda}, Keiichi and {Ray}, Alak K. and {Nayana}, A.~J.},
        title = "{Chandra's Insights into SN 2023ixf}",
      journal = {\apjl},
         year = 2024,
        month = mar,
       volume = {963},
       number = {1},
          eid = {L4},
        pages = {L4},
          doi = {10.3847/2041-8213/ad275d},
archivePrefix = {arXiv},
       eprint = {2311.04384},
 primaryClass = {astro-ph.HE},
       adsurl = {https://ui.adsabs.harvard.edu/abs/2024ApJ...963L...4C}
}

@ARTICLE{jacobsongalan25a,
       author = {{Jacobson-Gal{\'a}n}, Wynn},
        title = "{SN 2023ixf: The Closest Supernova of the Decade}",
      journal = {Universe},
         year = 2025,
        month = jul,
       volume = {11},
       number = {7},
          eid = {231},
        pages = {231},
          doi = {10.3390/universe11070231},
archivePrefix = {arXiv},
       eprint = {2507.08078},
 primaryClass = {astro-ph.HE},
       adsurl = {https://ui.adsabs.harvard.edu/abs/2025Univ...11..231J}
}

@article{Koribalski_2004,
doi = {10.1086/421744},
url = {https://doi.org/10.1086/421744},
year = {2004},
month = {jul},
publisher = {},
volume = {128},
number = {1},
pages = {16},
author = {Koribalski, B. S. and Staveley-Smith, L. and Kilborn, V. A. and Ryder, S. D. and Kraan-Korteweg, R. C. and Ryan-Weber, E. V. and Ekers, R. D. and Jerjen, H. and Henning, P. A. and Putman, M. E. and Zwaan, M. A. and de Blok, W. J. G. and Calabretta, M. R. and Disney, M. J. and Minchin, R. F. and Bhathal, R. and Boyce, P. J. and Drinkwater, M. J. and Freeman, K. C. and Gibson, B. K. and Green, A. J. and Haynes, R. F. and Juraszek, S. and Kesteven, M. J. and Knezek, P. M. and Mader, S. and Marquarding, M. and Meyer, M. and Mould, J. R. and Oosterloo, T. and O’Brien, J. and Price, R. M. and Sadler, E. M. and Schröder, A. and Stewart, I. M. and Stootman, F. and Waugh, M. and Warren, B. E. and Webster, R. L. and Wright, A. E.},
title = {The 1000 Brightest HIPASS Galaxies: H I Properties},
journal = {The Astronomical Journal}
}

@ARTICLE{jacobsongalanetal25b,
       author = {{Jacobson-Gal{\'a}n}, W.~V. and {Dessart}, L. and {Kilpatrick}, C.~D. and {Patel}, P.~J. and {Auchettl}, K. and {Tinyanont}, S. and {Margutti}, R. and {Dwarkadas}, V.~V. and {Bostroem}, K.~A. and {Chornock}, R. and {Foley}, R.~J. and {Abunemeh}, H. and {Ahumada}, T. and {Arunachalam}, P. and {Bustamante-Rosell}, M.~J. and {Coulter}, D.~A. and {Gall}, C. and {Gao}, H. and {Guo}, X. and {Jones}, D.~O. and {Hjorth}, J. and {Kaewmookda}, M. and {Kasliwal}, M.~M. and {Kaur}, R. and {Larison}, C. and {LeBaron}, N. and {Miao}, H.-Y. and {Narayan}, G. and {Pan}, Y.-C. and {Park}, S.~H. and {Patra}, K.~C. and {Qin}, Y. and {Ransome}, C.~L. and {Rest}, A. and {Rho}, J. and {Rose}, S. and {Sears}, H. and {Swift}, J.~J. and {Taggart}, K. and {Villar}, V.~A. and {Wang}, Q. and {Zenati}, Y. and {Zhou}, H.},
        title = "{A Panchromatic View of Late-time Shock Power in the Type II Supernova 2023ixf}",
      journal = {\apjl},
         year = 2025,
        month = nov,
       volume = {994},
       number = {1},
          eid = {L14},
        pages = {L14},
          doi = {10.3847/2041-8213/ae157a},
archivePrefix = {arXiv},
       eprint = {2508.11747},
 primaryClass = {astro-ph.HE},
       adsurl = {https://ui.adsabs.harvard.edu/abs/2025ApJ...994L..14J}
}
\bibliographystyle{aasjournalv7}



\end{document}